\documentclass[
 reprint,
 amsmath,amssymb,
 aps,
 prx,
 longbibliography,a4paper, amsfonts, amssymb, reprint, superscriptaddress,showkeys,aps, physrev, nofootinbib,twoside,twocolumn
]{revtex4-2}
\usepackage{tikz}
\usetikzlibrary{matrix, fit}
\usepackage{natbib}
\usepackage{graphicx}
\usepackage{dcolumn}
\usepackage{bm}
\usepackage{float}
\usepackage{placeins}
\usepackage{listings}
\usepackage{xcolor}
\usepackage{braket}
\usepackage{amsmath}
\usepackage{MnSymbol}
\usepackage{svg}
\usepackage{textcomp}
\usepackage{mathtools}
\usepackage[normalem]{ulem}
\usepackage{hyperref}
\usepackage{silence}
\usepackage{booktabs}
\usepackage{multirow}
\usepackage[normalem]{ulem}
\graphicspath{{./pics/}, [interpolate,keepaspectratio]}
\usepackage{subfigure}

\usepackage{graphicx} % 优先加载
\usepackage{algorithm} % 算法宏包
\usepackage{algpseudocode} % 算法伪代码
\usepackage[font=small]{caption} % 标题格式控制

\usepackage{multirow}

\begin{document}
\title{Self-attention U-Net decoder for toric codes}

\author{Wei-Wei Zhang}
\email{wei-wei.zhang@nwpu.edu.cn}
\affiliation{%
School of Computer Science, Northwestern Polytechnical University, Xi’an 710129, China 
}

\author{Zhuo Xia}
\affiliation{School of Computer Science, Shanghai Jiao Tong University, Shanghai, 200240, China.}
\affiliation{%
School of Computer Science, Northwestern Polytechnical University, Xi’an 710129, China 
}

\author{Wei Zhao}
\affiliation{
National Key Laboratory of Security Communication,  Chengdu 610041, China 
}
\author{Wei Pan}
\affiliation{%
School of Computer Science, Northwestern Polytechnical University, Xi’an 710129, China 
}

\author{Haobin Shi}
\affiliation{%
School of Computer Science, Northwestern Polytechnical University, Xi’an 710129, China 
}

\date{\today} % Leave empty to omit a date

\begin{abstract}
In the NISQ era, one of the most important bottlenecks for the realization of universal quantum computation is error correction. Stabiliser code is the most recognizable type of quantum error correction code. A scalable efficient decoder is most desired for the application of the quantum error correction codes. In this work, we propose a self-attention U-Net quantum decoder (SU-NetQD) for toric code, which outperforms the minimum weight perfect matching decoder, especially in the circuit level noise environments. Specifically, with our SU-NetQD, we achieve lower logical error rates compared with MWPM  and discover an increased trend of code threshold as the increase of noise bias. We obtain a high threshold of 0.231 for the extremely biased noise environment. The combination of low-level decoder and high-level decoder is the key innovation for the high accuracy of our decoder. With transfer learning mechanics, our decoder is scalable for cases with different code distances. Our decoder provides a practical tool for quantum noise analysis and promotes the practicality of quantum error correction codes and quantum computing.
\end{abstract}

%\keywords{first keyword, second keyword, third keyword}

\maketitle

\section{Introduction} \label{sec:introduction}
Quantum computing \cite{nielsen_chuang_2019} has attracted significant attention due to its potential to address challenges that classical computing faces with remarkable efficiency. 
With the entanglement and superposition properties, quantum computers possess parallel computing capabilities that classical computers lack and have demonstrated their advantage in various fields including 
cryptography \cite{shor1999polynomial, gisin2002quantum}, database searches \cite{grover1996fast, ambainis2007quantum,zhang2024quantum}, combinatorial optimization \cite{han2002quantum, farhi2014quantum}, molecular dynamics simulations \cite{peruzzo2014variational, romero2018strategies}, and machine learning \cite{havlivcek2019supervised, cong2019quantum, cerezo2021variational}. 
%In classical computing, a bit is restricted to a state of either 0 or 1 at any given time; conversely, in quantum computing, a qubit may represent any unit vector within the two-dimensional complex vector space, known as Hilbert space. Typically, a quantum state is denoted as $|\psi\rangle = a |0\rangle + b |1\rangle$, where $a, b \in \mathbb{C}$ and $|a|^2 + |b|^2 = 1$. Quantum operations on qubits are reversible and can be represented by unitary matrices. 

In the NISQ era, the bottleneck for the development of large-scale quantum computation is error correction \cite{preskill2018quantum}. Despite the realisation of the 1000 qubit milestone \cite{castelvecchi2023ibm}, the qubits exhibit high error rates ranging from $10^{-3}$ to $10^{-2}$, exceeding the threshold needed for practical applications (below $10^{-10}$). Quantum error correction is the milestone to be achieved for the resistance to noise before the realization of universal quantum computation \cite{shor1996fault}. The threshold theorem suggests that dispersing a single logical qubit's information across multiple physical qubits through redundant encoding can facilitate error correction with arbitrary precision \cite{aharonov1997fault}. Surface code is one of the most popular and hardware-friendly error-correcting codes \cite{fowler2012surface}. Traditional surface code decoding algorithms, such as Minimum Weight Perfect Matching (MWPM) \cite{edmonds1965paths}, Union Find (UF) \cite{huang2020fault} and Tensor Network (TN) \cite{bravyi2014efficient} decoders, encounter some practical issues including scalability and accuracy. Recent advancements in quantum error correction decoders include data-driven quantum error correction strategies employing neural network frameworks like multi-layer perceptrons (MLP) \cite{chamberland2018deep}, convolutional neural networks (CNN) \cite{breuckmann2018scalable}, and transformer \cite{wang2023transformerqecquantumerror,bausch2024learning}, 
have shown efficiency in addressing these challenges, and scalable and rapid decoding solutions have been developed for extensive surface codes \cite{gicev2023scalable}.

Nevertheless, the machine-learning decoders encounter specific restrictions. First, fixed input and output dimensions in models like MLP necessitate extensive retraining for adjustments in code distance, adding significant overhead. Besides, Recent studies \cite{fu2023benchmarkingmachinelearning} highlight the critical role of models' capacity to capture long-range dependencies in QEC, a task at which CNN-based decoders often falter because of the limitation from their smaller receptive fields. Transformers are effective at capturing global dependencies but do not inherently consider the translational invariance crucial to surface codes~\cite{vaswani2017attention, alammar2018illustrated}. The training of transformers often requires much larger datasets compared with other not-so-large AI models, which is a significant drawback given the data scarcity in quantum error correction scenarios. Additionally, Transformers are computationally intensive, leading to significant overhead in both training and inference. Furthermore, when representing surface codes in two-dimensional graphical formats, existing models fail to consider unique structural characteristics like the toroidal nature of surface codes and the connection between error chains and the lattice structures, leading to pronounced edge effects. Additionally, most study on surface code decoders focuses on low-level decoders for locating errors in data qubits, overshadowing the need for high-level decoders that address logical errors.

In this work,  We introduce a Self-attention U-Net  decoder (SU-NetD) for toric codes, a multilevel decoder leveraging U-Net and multi-head self-attention mechanisms to relieve the restrictions mentioned above and obtain higher decoder accuracy. This framework includes a low-level decoder tasked with detecting and correcting errors in data qubits and a high-level decoder designed to rectify any logical errors that may arise from the low-level decoder's processes. Additionally, our high-level decoder can be integrated into the post-processing stages of classical decoders, thereby markedly enhancing their decoding capabilities. In summary, the main contributions of our work can be summarized as follows:
\begin{itemize}
  \item We present an innovative multilevel neural network decoder architecture tailored for toric codes. In analogy to the semantic segmentation task in computer vision, our low-level decoder processes syndromes to predict the recovery chain operations, and furthermore, the high-level decoder utilizes these predicted recovery chains alongside the original syndromes to identify and address logical errors introduced by the low-level decoder. Meanwhile, the integration of our high-level decoder into the MWPM decoder can also significantly improve its performance. 
  \item To address the unique toroidal structure and translational invariance of surface codes, we introduce circular padding and random translations to augment data. This approach is thoroughly integrated into our neural network architecture, significantly enhancing the decoder's accuracy, particularly in handling edge errors.
  \item To dramatically reduce training time and enable scalability to QEC with various code distances without retraining, 
  our model integrates transfer learning mechanics, and it works well, inspired by the work in Ref.\cite{wang2023transformerqecquantumerror}.
  \item We demonstrate that our decoder consistently outperforms the MWPM decoder under various physical errors and code distance scenarios. Especially, in the circuit level noise condition, our SU-NetD archives much lower logical error rates compared with MWPM, which is challenging for other neural network based decoders. 
\end{itemize}

The remainder of this paper is structured as follows: Section \ref{sec:backgroud and related works} reviews the background and challenges of quantum error correction, summarizes the principal characteristics of toric Code, and discusses the merits and limitations of various decoders designed for these codes. Section \ref{sec:methodology} presents our SU-NetQD framework's decoding process, including data enhancement strategies for handling toric codes' unique toroidal structure and translational invariance, and our transfer learning approach for accommodating different code distances. Section \ref{sec:experiments} presents our model's primary outcomes for various code distances and noise models, including the significant enhancements achieved in MWPM decoder performance through our high-level decoder and the ablation experiments considering the self-attention and transfer learning mechanics to prove the scalability and practicality of our method. Section \ref{sec:conclusions} provides a summary of our work.

\section{Background and Related Works} \label{sec:backgroud and related works}
\subsection{Quantum Error Correction}
Quantum computing has shown its quantum supremacy over traditional computing \cite{Harrow2017Quantum, arute2019quantum, zhong2020quantum}, yet the realization of large-scale universal quantum computers faces significant hurdles due to current hardware limitations. Ideally, within a closed quantum system, the computational power of a quantum computer would exponentially increase with the increase of the number of qubits, benefiting from their high levels of entanglement and superposition. These idealized qubits are often called logical qubits. However, no quantum system is entirely isolated and every qubit interacts with its surroundings to some extent, leading to quantum system noise. This noise progressively erodes the information stored within quantum states, obstructing the operations within quantum computers—a process known as quantum decoherence \cite{zurek2003decoherence}.

Quantum decoherence arises from multiple sources, such as temperature fluctuations, electromagnetic disturbances, quantum gate defects, and environmental interactions \cite{gill2024quantum}. Quantum errors, unlike their classical counterparts, manifest in more complex ways: phase noise changes the relative phase among a qubit's base states, while amplitude noise influences the probabilities of measuring distinct quantum states. Uncontrolled, these influences would transform an original quantum superposition into a classical mixed state, thereby erasing the distinctive quantum coherence and interference effects. This degradation makes computations futile and confines the scope and intricacy of feasible quantum algorithms. Quantum error correction employs entanglement to redundantly encode information across multiple qubits, enhancing the system's resilience to noise. Theoretically, if the physical error rate $p$ is maintained below the code's threshold in the corresponding type of noise, the rate of logical errors will drop exponentially with the increase of the code size\cite{steane1996multiple}. Therefore, increasing the redundancy in the quantum error correction codes makes it possible to achieve practical quantum computing applications.

\begin{figure*}[t]
    \centering
{{\subfigure(a)}\includegraphics[width=0.85\columnwidth]{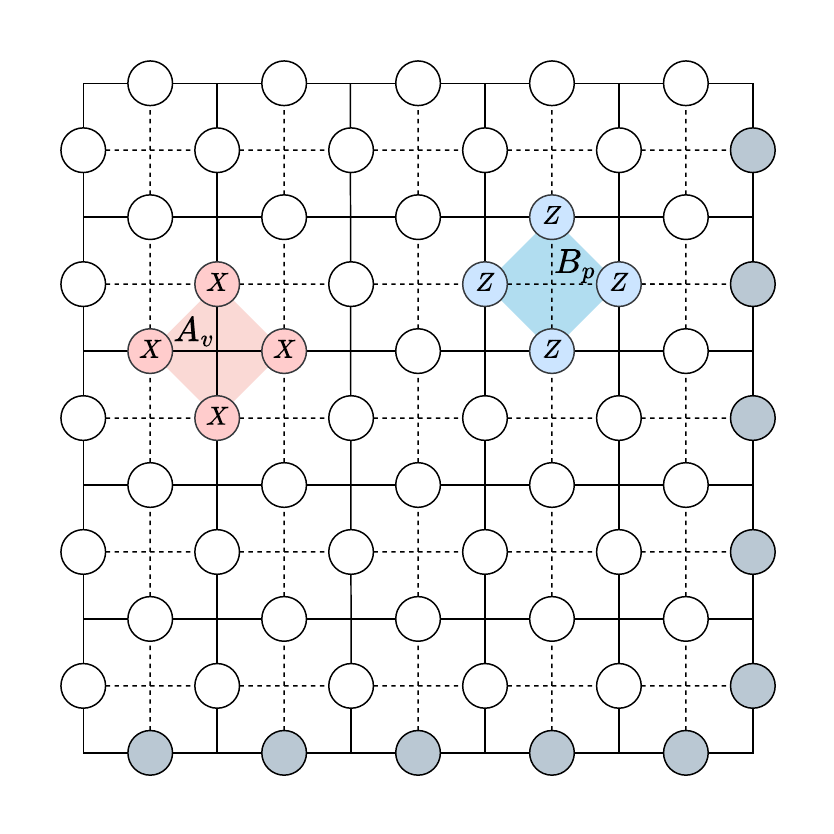}}
{{\subfigure(b)}
\includegraphics[width=0.9\columnwidth]{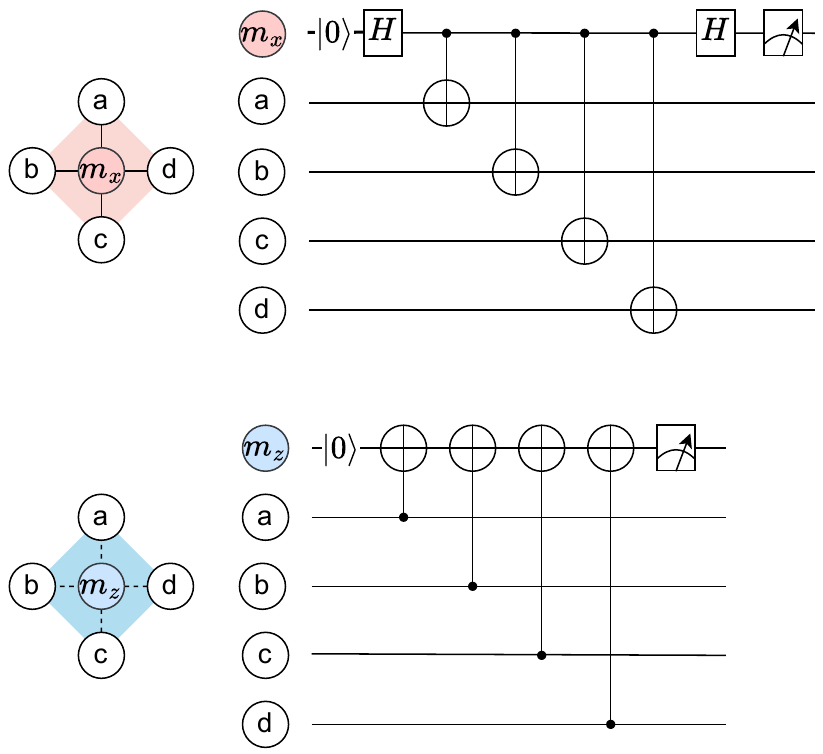}}
       \caption{(a) The structure of toric code and its stabilizer operators. The toric code is a square lattice with periodic boundary conditions, where each edge holds a qubit (represented by a circle). Solid lines represent the original lattice, and dashed lines represent the dual lattice. \(A_v\) and \(B_p\) are the toric code generators composed of tensor products of the Pauli operators \(X\) (red) or \(Z\) (blue), centered on the vertices and plaquettes of the original lattice, respectively. (b) Syndrome Measurement Circuit for toric code. Top: circuit for Z-type measurement. Bottom: circuit for X-type measurement.}
    \label{fig:1}
\end{figure*}

One of the most popular Quantum error correction codes is defined with stabilizer generators \cite{poulin2005stabilizer}. The generators commute with one another, and define a code-space by the intersection of their +1 eigenspaces. 
%A stabilizer state for \(n\) qubits is the common +1 eigenstate of a set of operators that form a commuting subset of a Pauli group, meaning all operators in the set mutually commute. 
In the presence of noise, identifying and correcting errors involves detecting the stabilizers' eigenvalues and applying recovery operations based on these detections, aiming to restore the quantum state to the logical codespace. Different decoding strategies may be required to handle various types of noise effectively, highlighting the need for tailored approaches in quantum error correction to maximize the potential of quantum computing \cite{terhal2015quantum}.

\subsection{Toric Code}
In the current landscape of quantum hardware, qubits are confined by the physical geometry of the devices, which typically restricts multi-qubit quantum gate operations to only those physically adjacent qubits. Operations on qubits that are spatially separated often necessitate the inclusion of intermediary qubits and the breakdown of these operations into multiple local ones \cite{mottonen12006decompositions}, incurring significant costs and noise. In this context, a class of encoding schemes known as surface codes \cite{dennis2002topological} has emerged as a viable option for achieving fault-tolerant quantum computing. These codes exhibit robust resistance to local perturbations and align well with the geometric configurations of contemporary quantum computers. The toric code \cite{kitaev2003fault}, introduced by A. Kitaev in 1997, is a prominent example of such codes, forming a fascinating bridge between topology and quantum computing.

In its seminal formulation, the toric code was designed for quantum circuits arranged on a toroidal surface. Specifically, it involves a two-dimensional square lattice endowed with periodic boundary conditions, wherein a physical qubit is positioned on each edge of the lattice, as depicted in Fig.~\ref{fig:1}(a). This arrangement leverages the topology of the torus to facilitate error correction across the lattice, exemplifying the integration of physical constraints and quantum coding principles.

\begin{figure*}[!ht]
    \centering
    % \vspace{0.30cm} %设置与上面正文的距离`
    %\setlength{\abovecaptionskip}{0.0cm}   %调整图片标题与图距离`
    %\setlength{\belowcaptionskip}{0.0cm} %调整图片标题与下文距离
    \includegraphics[width=0.8\linewidth]{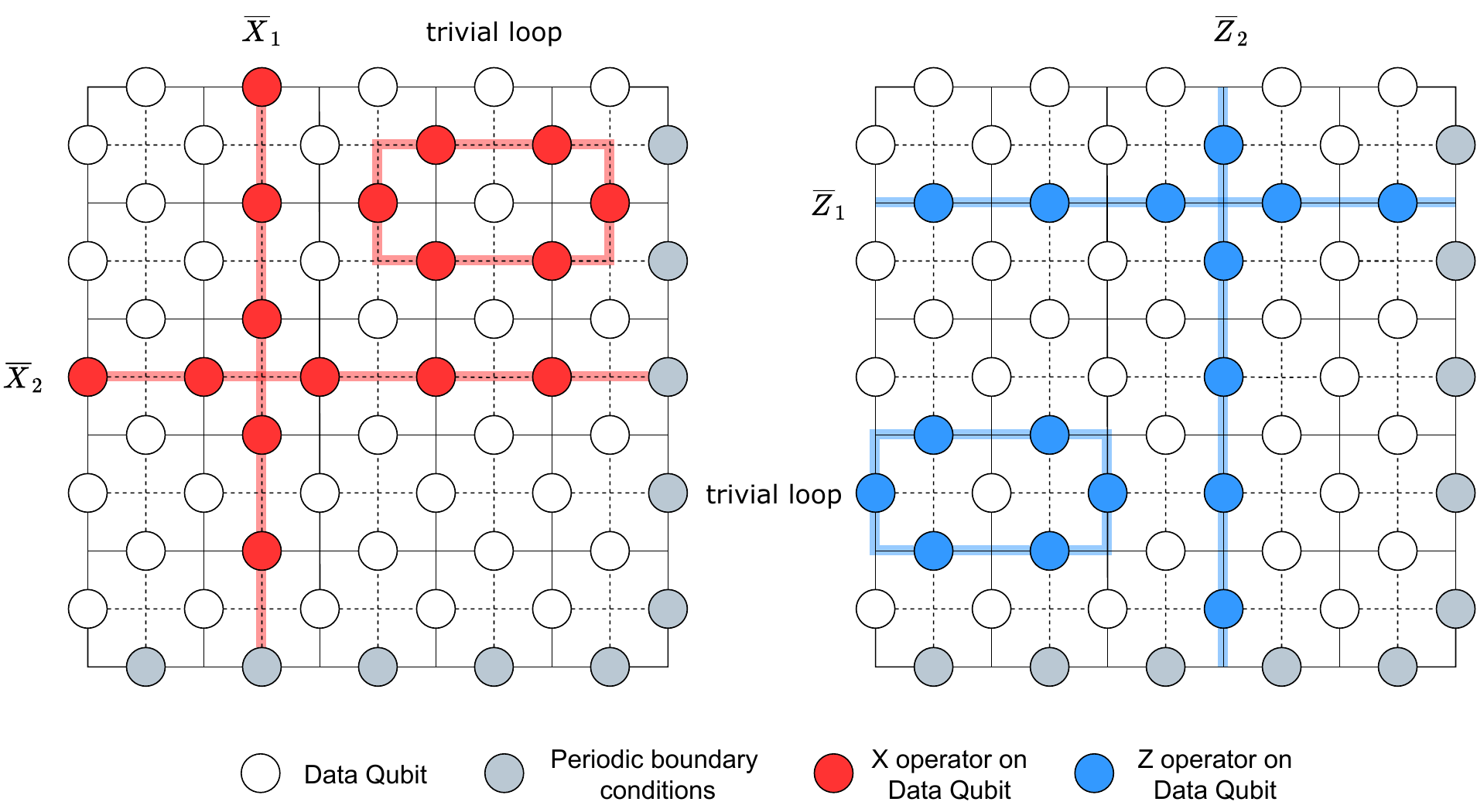}
    \caption{The illustration of toric code. The logical operators are global loops and cannot be shrunk to a point through continuous deformations. The local loops are trivial and can be shrunk to a single vertex or removed entirely by local operations. }
    \label{fig:2}
\end{figure*}

In this model, the logical qubits' quantum states are defined by the ground state of the Hamiltonian \( H \), expressed as:
\begin{equation}
    H = -\sum_v A_v - \sum_p B_p
\end{equation}
where the first sum is over all vertices \(v\) and the second is over all faces or plaquettes \(p\). The operators \( A_v \) and \( B_p \) are tensor products of the Pauli operators \( X \) and \( Z \) acting on individual qubits, forming the stabilizers of the toric code:
\begin{equation}
    A_v = \prod_{j \in \text{star}(v)} X_j, \qquad B_p = \prod_{j \in \text{boundary}(p)} Z_j
\end{equation}
where \( \text{star}(v) \) is the set of edges adjacent to vertex \( v \), and \( \text{boundary}(p) \) is the set of edges forming the boundary of plaquettes \( p \).

\begin{figure*}[!ht]
    \centering
    \includegraphics[width=0.95\linewidth]{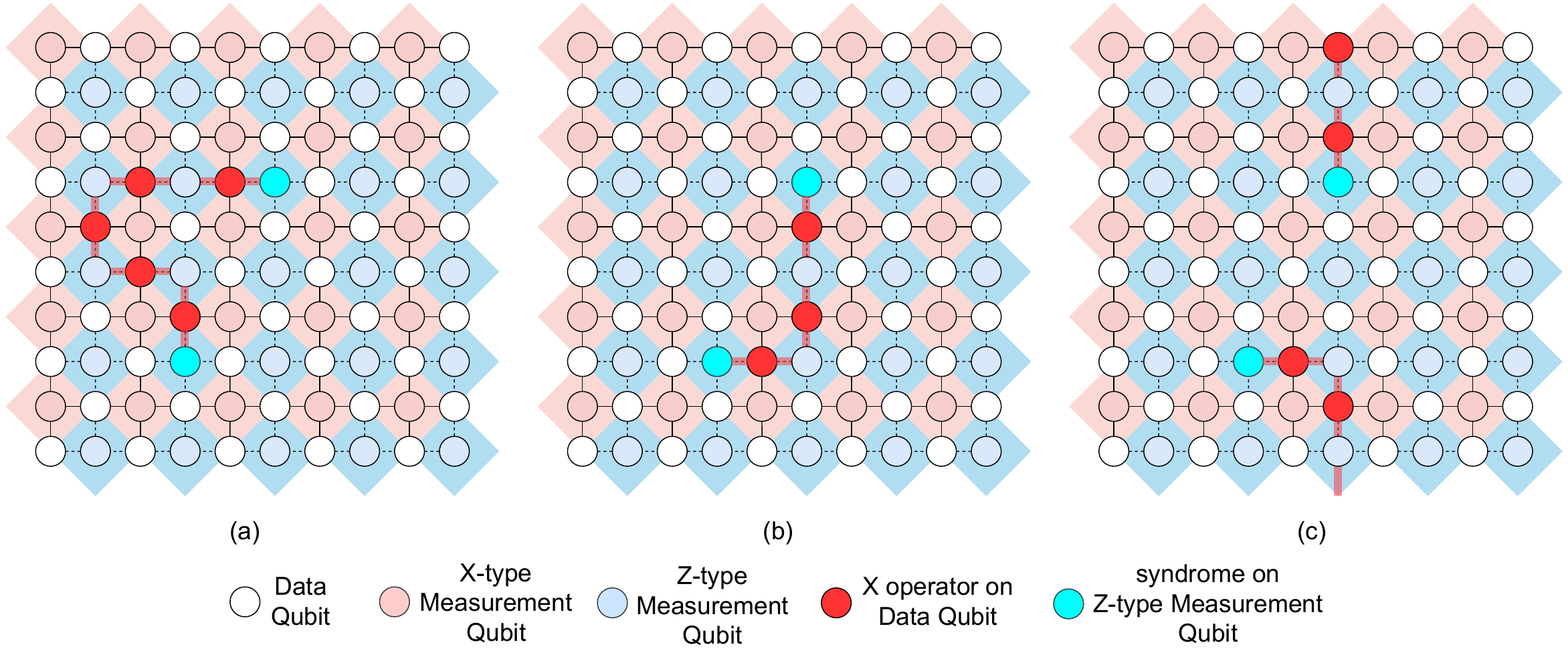}
    \caption{An
    example of (a) QEC errors, (b) the correct recovery chain,  (c) the incorrect recovery chain
    Errors (left). A non-closed error chain causes the eigenvalues of the stabilizers at its ends to become -1, which results in the measurement outcomes on the corresponding measurement qubits becoming $|1\rangle$ as shown in (a). The quantum decoder needs to infer the recovery chain operations from the syndrome to eliminate it, but incorrect inference can lead to logical errors as shown in (b) and (c).}
    \label{fig:3}
\end{figure*}
Assuming the square lattice's side length is \( d \), the toric code comprises \( 2d^2 \) physical qubits and \( 2d^2 - 2 \) linearly independent stabilizer operators. The ground state of \( H \) is the common eigenstate where all stabilizers' eigenvalues are \( +1 \). Hence, a toric code's corresponding Hilbert space dimension is 4, indicating a system degeneracy of 4, equivalent to encoding two logical qubits.

The toric code's four orthogonal basis states are \( |00\rangle \), \( |10\rangle \), \( |01\rangle \), and \( |11\rangle \). Operations forming closed loop sequences by combining several stabilizer operators, as illustrated in Fig.~\ref{fig:2}, are termed trivial loops because they do not alter the original logical quantum state. Conversely, logical operators \( \bar{X}_1 (\bar{X}_2) \) and \( \bar{Z}_1 (\bar{Z}_2) \), which involve continuous \( X \) and \( Z \) operations along specific orientations of the toric code's lattice, effect transitions such as \( |00\rangle \rightarrow |10\rangle \) and \( |10\rangle \rightarrow -|10\rangle \), respectively. These operations, not representable by stabilizer operator products, are known as non-trivial loops.

In toric codes, physical qubit errors typically manifest as sequences of global error chains, transitioning the system from its ground state to an excited state, thus disrupting the encoded state. 

Although the quantum error is continuous, it can be decomposed into a linear combination of Pauli errors~\cite{shor1995scheme}. If a decoder can correct $X$ and $Z$ errors, it is equipped to address any unitary error affecting the system \cite{knill1997theory}. Specifically, we define error chain operators as follows:
\begin{equation}
    S^x(t)=\prod_{j\in t} X_j, \quad S^z(t')=\prod_{j\in t'} Z_j
\end{equation}
where \(t\) and \(t'\) represent chains on the dual and original lattices, respectively. The objective of quantum error correction is to restore the system to its ground state without  performing operations that alter the state of the original logical quantum bits. A non-closed \(X\) error chain \(S^x(t)\) will result in a pair of anyons (magnetic vortices) on the plaquettes at its ends, flipping the eigenvalues of the stabilizer \(B_p\) to -1. Similarly, a non-closed \(Z\) error chain \(S^z(t')\) generates a pair of anyons (electric charges) on the vertices at its ends, affecting the stabilizer \(A_v\)'s eigenvalues to -1. A \(Y = iXZ\) error chain leads to both types of anyons on the vertices and plaquettes at the ends.

A major challenge in quantum error correction arises in measurements, as it inevitably causes the collapse of the original quantum state. However, the stabilizers' eigenvalues can be measured without disturbing the logical qubits. As illustrated in Fig.~\ref{fig:1}(b), \(a, b, c, d\) are data qubits forming the logical quantum bits, with \(m_x\) and \(m_z\) as auxiliary measurement qubits placed at the vertices of the original and dual lattices. When \(A_v\) (\(B_p\)) has eigenvalues of +1, \(m_x\) (\(m_z\)) shows a \(|0\rangle\) result; if -1, a \(|1\rangle\) result is observed. These results collectively form the syndrome of the toric code under the given error.

Since different errors in toric code may yield the same syndrome, the error-to-syndrome map is a complicated process. Considering the syndromes from an \(X\) error chain (refer to Fig.~\ref{fig:3}), the decoder aims to provide recovery chain operations to cancel out these syndromes. If the recovery operations and the error form a trivial loop, the system returns to the original logical quantum state.  In the case that the error and the recovery chain form a non-trivial loop around the torus, the recovery operation results in a logical error. While the syndrome is removed, the decoding fails. 

In our work, we design a decoder composed of a low-level decoder and a high-level decoder, aiming to eradicate syndromes. The low-level decoder focuses on detecting and correcting data qubit errors, and the high-level decoder aims to recognize the logical errors introduced by the low-level decoder inspired by Ref.~\cite{varsamopoulos2019comparing}.

\begin{figure*}[!ht]
    \centering
    \includegraphics[width=0.75\linewidth]{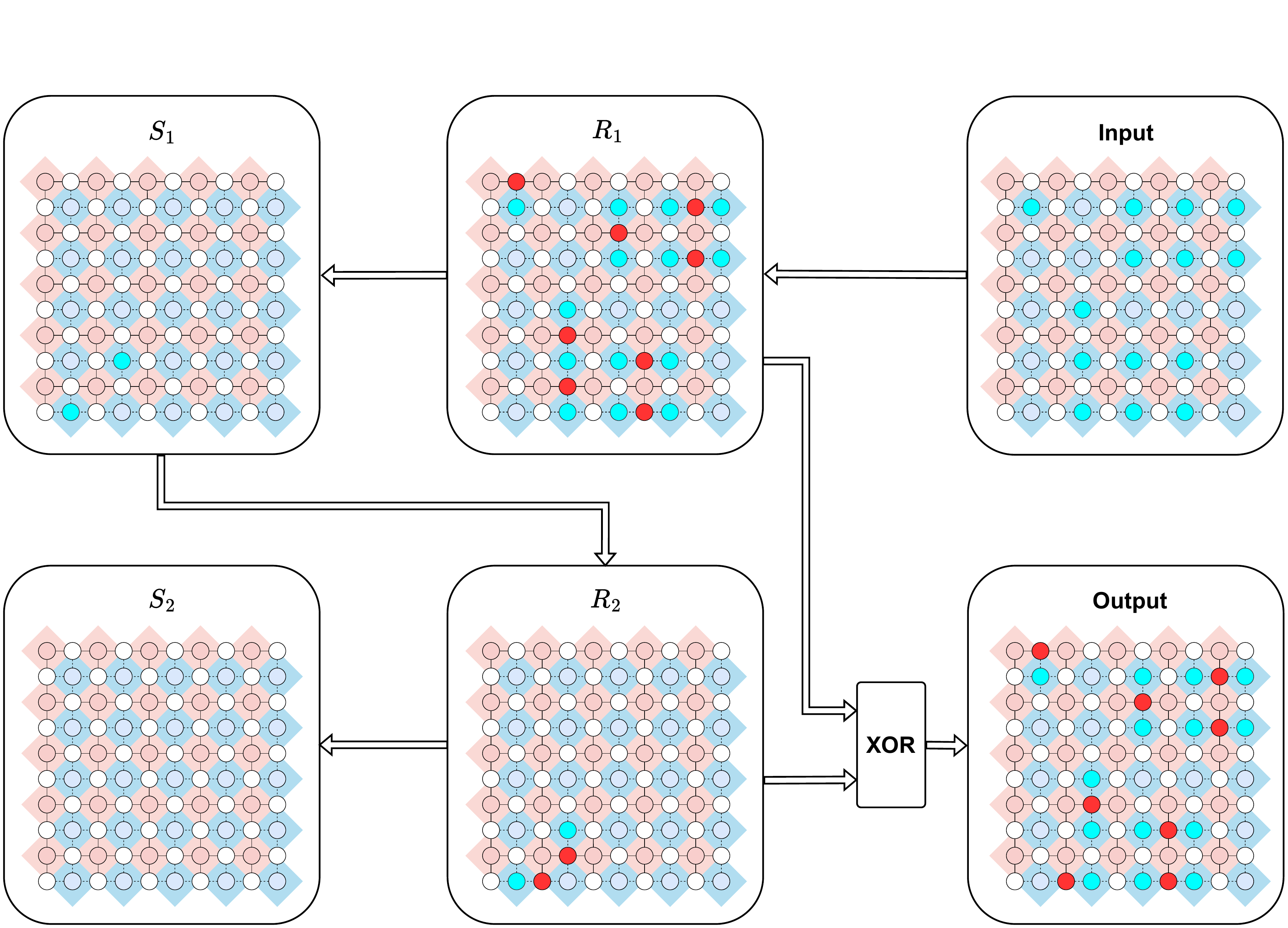}
    \caption{The scheme of our low-level decoder's two-round iterative decoding process. The input is the original syndrome. 
    The 1st round iteration provides a suggested recovery chain R1 and a syndrome S1 after the 1st round of error corrections.
    The 2nd round iteration provides a suggested recover chain R2 and the syndrome without errors S2.
    The output of the final suggested recovery chain is the xor of R1 and R2.
}
    \label{fig:4}
\end{figure*}

\subsection{Related Works}
The graph-based Minimum Weight Perfect Matching (MWPM) algorithm is the most widely adopted decoding algorithm. This algorithm seeks to identify the minimum number of errors that yield the same syndrome.  MWPM is renowned for its high error-correcting capabilities, especially beneficial at low error rates. However, its computational complexity, approximately \(O(N^3)\) \cite{kolmogorov2009blossom} with \(N\) representing the number of physical qubits and restricting its use in large quantum systems. The Union-Find (UF) decoder, despite its nearly linear time complexity~\cite{liyanage2023scalable}, occasionally exhibits suboptimal performance in error correction compared to the MWPM algorithm.

To address the issues mentioned above, machine learning, particularly neural network-based decoders, has gained attention recently for its advantageous properties. The Boltzmann machine initially demonstrated ML-based decoding in the toric code \cite{torlai2017neural}, which was succeeded by applications like multilayer perceptrons (MLP) \cite{chamberland2018deep} and Long Short-Term Memory Networks (LSTM) \cite{baireuther2018machine} in surface codes. Most modern data-driven QEC strategies now utilize sophisticated architectures like Convolutional neural networks (CNN) \cite{breuckmann2018scalable} and transformer \cite{wang2023transformerqecquantumerror,bausch2024learning}, 
which have shown effectiveness in quantum error correction. 
These trained neural network-based decoders achieve a theoretical time complexity of \(O(1)\). By considering the probability distributions of physical errors and the correlations between different types of errors,  these neural network-based decoders generally outperform traditional decoders (MWPM). But the training process of the transformer network requires an extensive data set, and the sequential setting is time-consuming compared to the ordinary neural network.

To further improve the performance of the decoder and its performance in a correlated noise environment, our work develops a multilevel decoder that incorporates a U-Net architecture with multi-head self-attention mechanisms~\cite{ronneberger2015u, siddique2021u}, aiming to innovate architecture design and improve scalability and practicality through multilevel decoding strategies.

\section{Self-attention U-Net decoder (SU-NetD) for toric codes} \label{sec:methodology}
In this section, we present our self-attention U-Net decoder for toric codes and introduce the problem mapping, the circulate padding and data augmentation methods,  and transfer learning mechanics techniques used there.  

\subsection{Overall workflow}
Our SU-NetD consists of a low-level decoder and a high-level decoder. The low-level decoder aims to take the extracted syndromes as input and predicts the recovery operations.  The high-level decoder aims to predict any logical errors the low-level decoder might have introduced by leveraging the recoveries and the original syndromes. 
The decoding process is deemed successful if the low-level decoder's recovery chains eliminate the syndromes and the high-level decoder's prediction of logical errors aligns with the combined effect of the recovery chains and the original errors.

Our low-level decoder introduces an iterative decoding strategy in the low-level decoding process. This iteration process aims to suppress the decoding failure rates resulting from the fact that while the error rate is high the recovery chain predicted directly by a low-level decoder may not eliminate the syndromes due to the degeneracy of syndromes. For our method  the syndrome measurement is only implemented for the preparation of the initial input for our method. There is no need for further syndrome measurement within our decoding process. Therefore, 
our low-level decoder does not rely on the internal simulator for intermediate syndromes. In each round of low-level decoding, the input consists of the residual syndromes of the previous round. The output recovery chain can further clear these syndromes. This process is repeated until the syndromes are entirely removed or a pre-set maximum number of iterations is reached. The final output is obtained by performing an XOR operation on the recovery chains predicted in each round.

Fig.~\ref{fig:4} illustrates an example of a two-round iterative decoding process with our low-level decoder. The input is the original syndromes. After the first round of decoding, a pair of syndromes remains uneliminated. This residual syndrome is then fed into the low-level decoder for a second round of decoding, during which all syndromes are eliminated. The final output is the XOR result of the recovery chains predicted in these two rounds. Furthermore, it can be observed that after the first round of decoding, the number of residual syndromes is significantly reduced, and their distribution resembles the result under low noise conditions. Therefore, to further improve the efficiency and accuracy of decoding, we can employ a decoder trained under low-noise conditions to decode these residual syndromes.

Both the output and the input of the low-level decoder are taken as the input of the high-level decoder, which indicates that the correlation between 
the recovery chain suggested by the low-level decoder and the syndromes is investigated by our high-level decoder. Different error chains might result in the same syndrome,  the key issue our high-level decoder aims to resolve and utilize to eliminate logical errors. Technically, our high-level decoder can be understood as a multi-label classification task.

\subsection{Problem mapping}
\begin{figure}[!ht]
    \centering
        \includegraphics[width=0.75\linewidth]{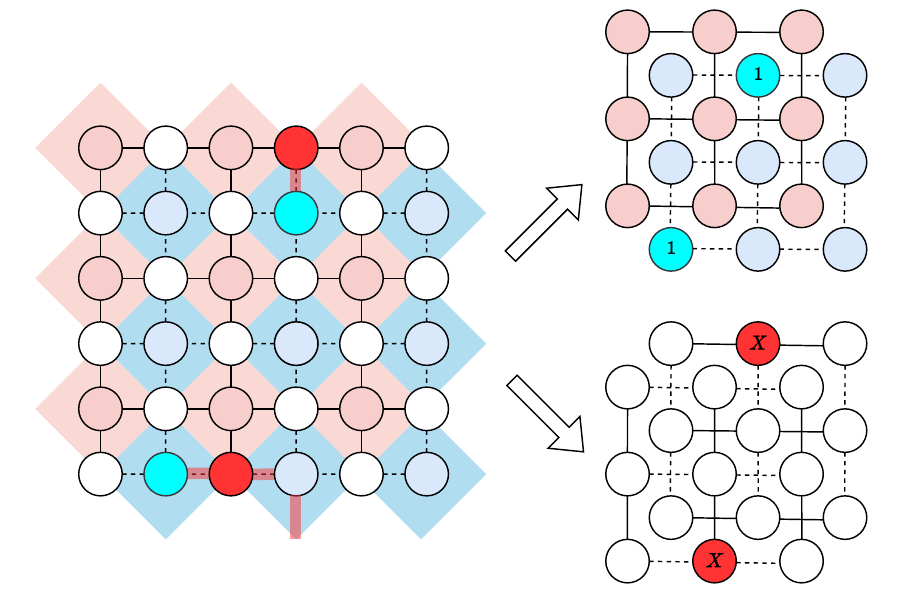}
    \caption{The syndromes and recoveries map of toric code under bit flip noise. The measurement results on the measurement qubits form the training dataset for the low-level decoder (top right), with the blue color noting the Z measurements and the pink color noting the X measurements, which are measured in sequential time steps; the recovery operations on the data qubits form the dataset labels for the low-level decoder (bottom right).}
    \label{fig:5}
\end{figure}

For a toric code with distance $d$, the results of measurement qubits on different lattices reflect different types of syndromes. We map the measurement qubits on the original and dual lattice vertices to two 2D structures of size $d \times d$, separately. Similarly, we map the data qubits to two 2D structures size of $d \times d$ according to their rows or columns in the original lattice.

As an example,  Fig.~\ref{fig:5} provides a 3$\times$3 toric code under bit flip noise, where only X errors occur on the data qubits. As shown in the top right of the figure, all Z-type measurement qubits (blue) form the first dimension of the channel, while all X-type measurement qubits (red) form the second dimension. The original syndromes can be represented as a three-dimensional tensor of size (2, d, d):
\begin{equation}
S_\text{origin} = \begin{bmatrix}
    \begin{bmatrix}
        0&1&0 \\
        0&0&0 \\
        1&0&0 
    \end{bmatrix}
    \begin{bmatrix}
        0&0&0 \\
        0&0&0 \\
        0&0&0
    \end{bmatrix}
    \end{bmatrix}
\label{eq:synd}
\end{equation}
Here, 0 and 1 indicate that the corresponding measurement qubit is in the quantum state \( |0\rangle \) or \( |1\rangle \), respectively, which correspond to eigenvalues of -1 or +1 of the respective stabilizer operator.

The recovery operations needed on the data qubits can be represented as a three-dimensional tensor of size (2, d, d):
\begin{equation}
Rec = \begin{bmatrix}
    \begin{bmatrix}
        0&0&0 \\
        0&0&0 \\
        0&1&0 
    \end{bmatrix}
    \begin{bmatrix}
        0&1&0 \\
        0&0&0 \\
        0&0&0
    \end{bmatrix}
    \end{bmatrix}
\end{equation}
Here, 0 and 1 represent \( I \) and \( X \) operations, respectively. 
As shown in the bottom right of Fig.~\ref{fig:5}, we encode the recovery operation on 
the data qubits in each column of the original lattice as the first dimension of the channel, and the operations on the data qubits in each row as the second dimension. 
Note that under other noise conditions, \( Z \) and \( Y \) operations might also be required, represented by 2 and 3, respectively.

\begin{figure*}[!ht]
    \centering
      \includegraphics[width=\linewidth]{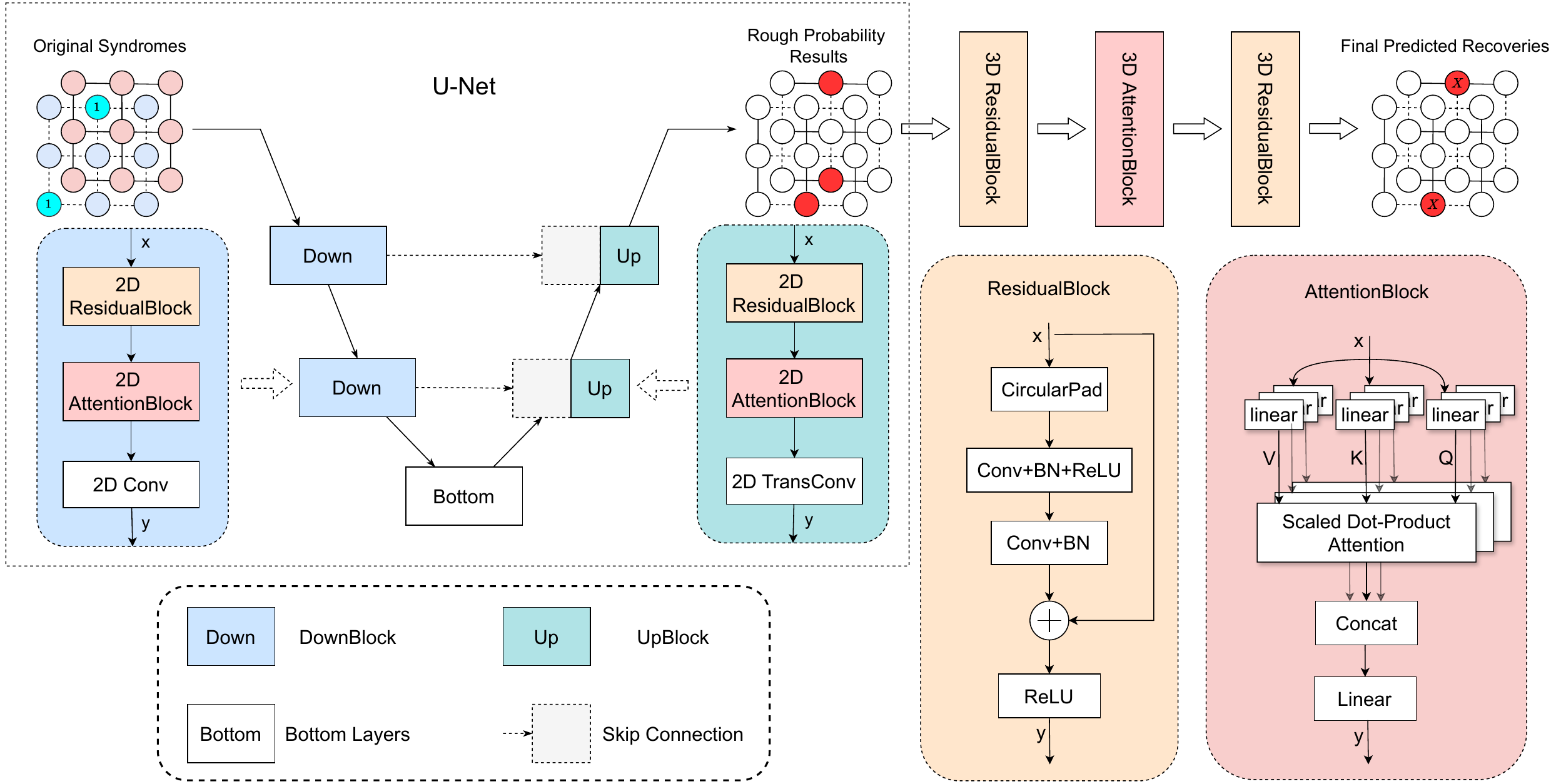}
    \caption{Illustration of SU-Net used in the low-level decoder. The model consists of a U-Net structure that extracts the intra-lattice information and a 3D residual block and attention block that fuses the inter-lattice information. The specific structures of the residual and attention blocks are shown in the lower right corner.}
    \label{fig:6.1}
\end{figure*}

In our machine learning mechanics, We take syndromes as training datasets and recovery chains as labels. Analogous to the semantic segmentation problem in computer vision, the low-level decoder in our SU-NetQD will predict the recovery operations on the data qubits of the same size by taking the corresponding error syndromes as input. The size of the output in the low-level decoder is $(4, 2, d, d)$, where the channel dimension represents the four different recovery candidate operations  (including $I, X, Y, Z$), and the value of each channel represents the confidence level of that category. 
The recovery candidate operation with the highest confidence is selected for the corresponding data qubit.

\begin{figure*}[!ht]
    \centering
 \includegraphics[width=0.75\linewidth]{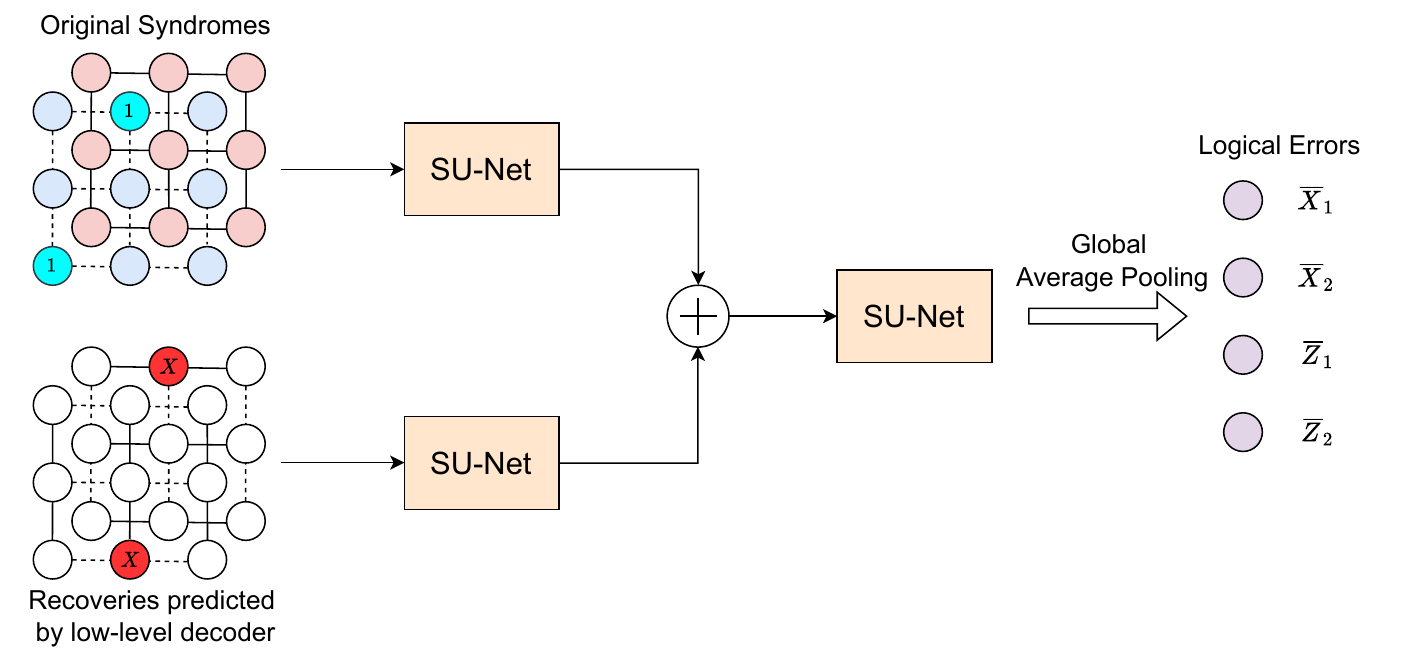}
    \caption{Illustration of the high-level decoder structure.}
    \label{fig:6.2}
\end{figure*}

\subsection{Circular padding and data augmentation}
In the problem mapping process, we map the error syndromes of the toric code to a 3D tensor representation. However, this approach loses the characteristic periodic boundary conditions of the toric code. For instance, in Fig.~\ref{fig:5}, the measurement qubits in the first row (column) are adjacent to those in the last row (column). But in the representation of Eq.~\ref{eq:synd} the distance between them appears as $d-1$, which does not reflect the boundary condition of toric codes. 
To enable the model's ability to handle boundary conditions, we use circular padding and data augmentation techniques.

For the data augmentation part, We perform random cyclic shifts on the training data by $\left\lfloor \frac{d}{2} \right\rfloor$ units in up, down, left, and right directions probabilistically, which can increase the number of boundary error samples. We use circular padding in the residual blocks of SU-Net instead of the classical zero padding, to preserve the periodic characteristics of the input data as shown in Fig.~\ref{fig:6.1}. Furthermore, we use a convolution kernel size of 3x3 following the circular padding to increase the receptive field for boundary syndromes and enhance the model's capability to capture boundary errors.

For the case shown in Fig.~\ref{fig:5}, after data augmentation by cyclic shifting 1 unit to the left, the syndrome in Eq.~\ref{eq:synd}  is converted to:
\begin{equation}
    S_{\mathrm{shift}} = \begin{bmatrix}
    \begin{bmatrix}
        1&0&0 \\
        0&0&0 \\
        0&0&1 
    \end{bmatrix}
    \begin{bmatrix}
        0&0&0 \\
        0&0&0 \\
        0&0&0
    \end{bmatrix}
    \end{bmatrix}
    \label{eq:synd_shift}
\end{equation}

The application of circular padding on the syndrome in Eq.~\ref{eq:synd_shift} converts the original syndrome into 
\begin{equation}
    S_{\mathrm{pad}} \! = \!\!\!
    \begin{bmatrix}
    \begin{tikzpicture}[baseline=(m-4-4.base)]
        \matrix (m) [matrix of math nodes,left delimiter={[\!\!\!\!},right delimiter={]}] {
            0 & 0 & 0 & 0 & 0 & 0 & 0 \!\!\!\! \\
            0 & 1 & 0 & 0 & 1 & 0 & 0 \!\!\!\! \\
            0 & 0 & 1 & 0 & 0 & 1 & 0 \!\!\!\! \\
            0 & 0 & 0 & 0 & 0 & 0 & 0 \!\!\!\! \\
            0 & 1 & 0 & 0 & 1 & 0 & 0 \!\!\!\! \\
            0 & 0 & 1 & 0 & 0 & 1 & 0 \!\!\!\! \\
            0 & 0 & 0 & 0 & 0 & 0 & 0 \!\!\!\! \\
        };
        \node[draw, dashed, inner sep=1pt, fit=(m-3-3)(m-5-5)] {};
    \end{tikzpicture}
    \begin{tikzpicture}[baseline=(n-4-4.base)]
        \matrix (n) [matrix of math nodes,left delimiter={[\!\!\!\!},right delimiter={]}] {
            0 & 0 & 0 & 0 & 0 & 0 & 0 \!\!\!\! \\
            0 & 0 & 0 & 0 & 0 & 0 & 0 \!\!\!\! \\
            0 & 0 & 0 & 0 & 0 & 0 & 0 \!\!\!\! \\
            0 & 0 & 0 & 0 & 0 & 0 & 0 \!\!\!\! \\
            0 & 0 & 0 & 0 & 0 & 0 & 0 \!\!\!\! \\
            0 & 0 & 0 & 0 & 0 & 0 & 0 \!\!\!\! \\
            0 & 0 & 0 & 0 & 0 & 0 & 0 \!\!\!\! \\
        };
        \node[draw, dashed, inner sep=1pt, fit=(n-3-3)(n-5-5)] {};
    \end{tikzpicture}
    \end{bmatrix}
\end{equation}

\subsection{Decoder design}
Our SU-NetQD is a novel SU-Net model based on a 2D U-Net structure combined with a self-attention mechanism, which consists of a low-level decoder and a high-level decoder. 
Our low-level decoder's architecture is depicted in Fig.~\ref{fig:6.1}. 
We use the U-Net structure to extract detailed lattice information from the input error syndromes. Subsequently, we apply the 3D residual blocks and attention blocks to integrate inter-lattice contextual information, refining the predictions and generating the final recoveries. Due to the small size of the test data, we use convolutional layers without padding for downsampling instead of traditional pooling layers. 
Both the downsampling and upsampling blocks of the U-Net use 2D residual blocks and attention blocks to fully extract local features and long-range dependencies within the lattice, as shown in the bottom right corner of Fig.~\ref{fig:6.1}. Each residual block applies circular padding with a padding number of 2, and uses two consecutive 3x3 convolutional kernels to extract local features. 
Subsequently, we apply batch normalization for data standardization. 
The residual connections merge the contextual information before and after convolution, enhancing the accuracy and efficiency of the results.

Our multi-head self-attention block applies multiple sets of distinct linear projections to transform the input into various queries (Q), keys (K), and values (V) sets. These transformed queries, keys, and value sets are fed parallelly into the attention aggregation layer, where the scaled dot-product attention mechanism is applied to compute the different heads. The obtained heads capturing various features are concatenated and processed through another learnable linear projection. This process facilitates the learning of long-range dependencies within the lattice.

Our low-level decoder utilizes the U-Net structure to effectively extract local features and aggregate information obtained from the contextual information in similar error syndromes and generates an initial prediction. 
However, it overlooks the correlation between $X$ and $Z$ errors and does not account for the spatial continuity between adjacent lattices, which limits its performance under complex noise conditions.  
Meanwhile, the 3D structure of the lattice also affects the 
 recovery operations on the data qubits.  
To achieve more refined predictions, we concatenate  3D residual and attention blocks after the 2D U-Net structure to capture and integrate 3D spatial contextual information. Our low-level decoder's output shape is $(4, 2, d, d)$, where the channel dimension represents four different recovery operations: $I, X, Y, Z$, each represented as one-hot code.

Our high-level decoder takes the original error syndromes,  which are used as the input of low-level decoder also, and the recovery chain provided  by the low-level decoder, i.e. the output of the low-level decoder, as input to predict the logical effects, where it requires no extra syndromes measurements required. The structure of our high-level decoder is shown in Fig.~\ref{fig:6.2}, which comprises of three SU-Net. 
At the front end of our high-level decoder, two SU-Nets are employed independently to extract intrinsic features from the original syndromes and the recovery chains simultaneously. The number of output channels in these two SU-Nets is modified to 1, resulting in an output shape of $(1, 2, d, d)$ for ease of subsequent processing.  The outputs of these two SU-Nets are combined to integrate the information from both sources, where the first dimension of the data is compressed. The resulting tensor is fed into the third SU-Net for final feature extraction, producing an output size of $(4, 2, d, d)$.  A global average pooling layer averages all elements within each channel, and the resulting values are used for classification. The values of the four channels represent the confidence levels of logical errors $\{\bar{X}_{1}, \bar{X}_{2}, \bar{Z}_{1}, \bar{Z}_{2}\}$ respectively.

A pseudocode description of our decoding procedure is presented in Algorithm \ref{pseudocode}.

\begin{algorithm}[H]
\renewcommand{\algorithmicrequire}{\textbf{Input:}}
\renewcommand{\algorithmicensure}{\textbf{Output:}}
\caption{SU-NetQD Decoding Algorithm}
\label{pseudocode}
\begin{algorithmic}[1]
\Require 
\Statex $S_{\text{origin}}$: initial syndromes of shape $(2, d, d)$
\Statex $N_{max}$: maximum number of iterations
\Statex $T$: confidence threshold for logical errors
\Ensure 
\Statex $R_{\text{final}}$: final predicted recoveries of shape $(2, d, d)$
\Statex $logical\_errors$: predicted logical errors

\State Initialize $i \gets 0$
\State $S_{\text{current}}  \gets S_{\text{origin}}$
\State $R_{\text{final}} \gets \mathbf{0}_{(2, d, d)}$

\While{$i < N_{max}$ and any($S_{\text{current}}$)}
    \State $low\_level\_output \gets \text{LowLevelDecoder}(S_{\text{current}})$
    \State $R_{\text{current}} \gets \text{argmax}(low\_level\_output, \text{axis}=0)$
    \State $S_{\text{current}} \gets \text{UpdateSyndromes}(S_{\text{current}}, R_{\text{current}})$
    \State $R_{\text{final}} \gets R_{\text{final}} \oplus R_{\text{current}}$
    \State $i \gets i + 1$
\EndWhile

\If{not any($S_{\text{current}}$)}
    \State $high\_level\_output \gets \text{HighLevelDecoder}(S_{\text{origin}}, R_{\text{final}})$
    \State $logical\_errors \gets [0, 0, 0, 0]$
    \For{$j \in \{1, 2, 3, 4\}$}
        \If{$high\_level\_output[j] > T$}
            \State $logical\_errors[j]=1$
        \EndIf
    \EndFor
\Else
    \State Decoding failed
\EndIf

\State \Return $R_{\text{final}}$, $logical\_errors$
\end{algorithmic}
\end{algorithm}

\subsection{Transfer learning}
Surface codes with different code distances demonstrate different tolerance to the same noisy environment. The longer the code distance is, the higher the noise tolerance ability the code has. 
Adapting decoding strategies to the same code with different distances requires extensive retraining, which is computationally intensive. Inspired by Ref.~\cite{wang2023transformerqecquantumerror}, which is the first to integrate transfer learning into a transformer-based quantum decoder. Our SU-NetQD uses a transfer learning approach to achieve high efficiency. This strategy harnesses the similarities in error distributions across codes with different distances under identical physical error conditions, by fine-tuning the pre-trained models on toric codes with new distances. 

The application of transfer learning significantly reduces the necessary training time. 
Empirical findings indicate that for toric codes with smaller code distances, the quality of the training dataset plays a more critical role in enhancing predictive accuracy than the complexity of the neural network itself. Applying transfer learning to adapt models trained on smaller code distances to tasks involving larger ones has been shown to be more efficient. This adaptation optimizes performance and underscores the practical benefits of transfer learning in managing quantum error correction efficiently.

\begin{figure*}[!t]
    \centering
    {{\subfigure(a)}
    \includegraphics[width=0.46\linewidth]{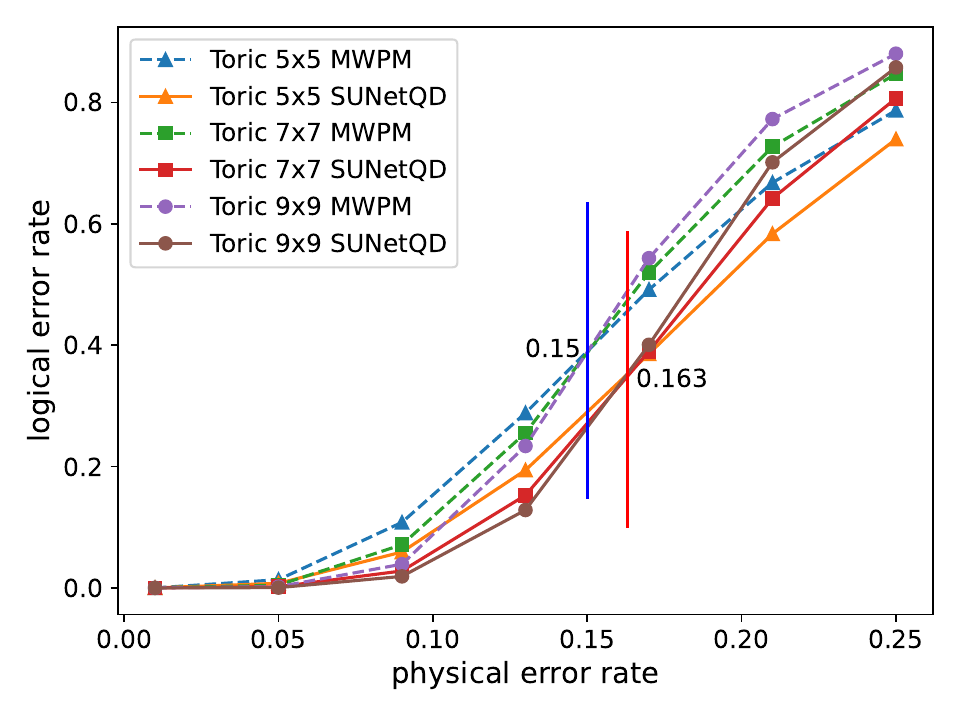} }
    {{\subfigure(b)}
     \includegraphics[width=0.46\linewidth]{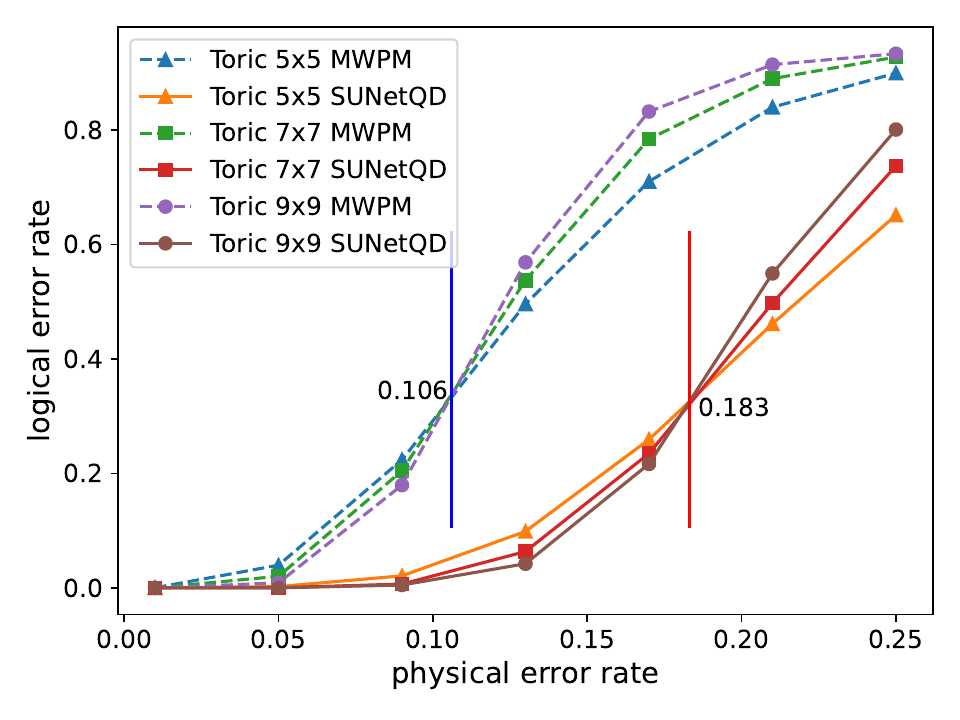}}
     {{\subfigure(c)}
     \includegraphics[width=0.46\linewidth]{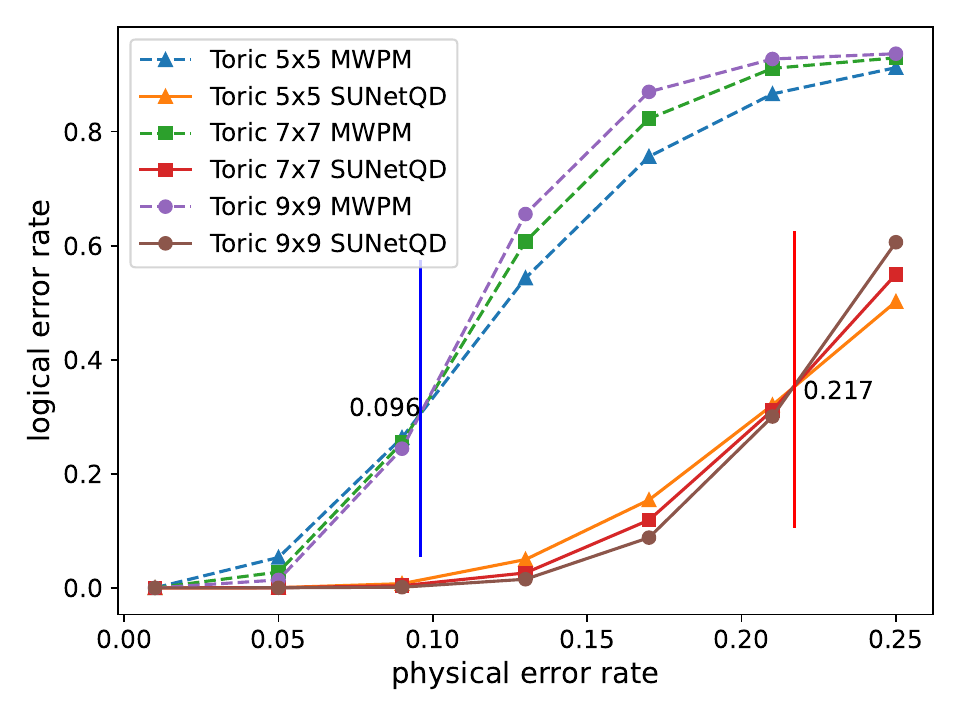} }
     {{\subfigure(d)}
   \includegraphics[width=0.46\linewidth]{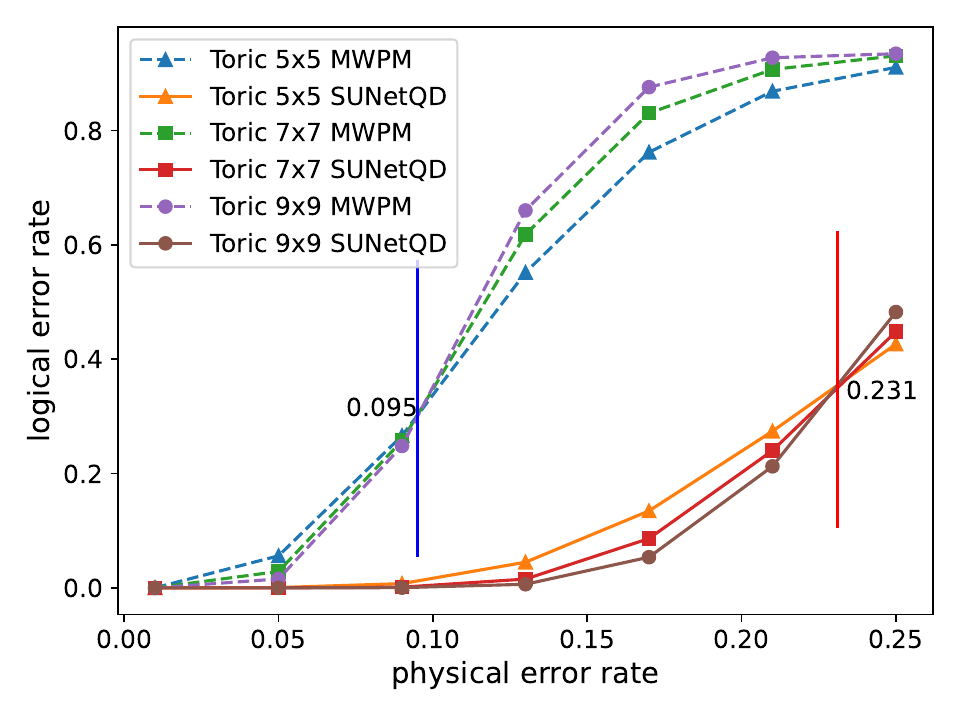} }
    \caption{Comparison of the decoding performance (logical error rate (LER)) of SU-NetQD and MWPM under various physical error rates of different noise models. 
   The X-axis represents the physical error rate and the Y-axis represents the corresponding logical error rate. The blue color vertical line locates the threshold revealed by MWPM and the red color vertical line locates the threshold revealed by our SU-NetQD.(a) depolarizing noise, 
   (b) biased depolarizing noise with $\eta=5$, 
   (c) biased depolarizing noise with $\eta=50$, 
   (d) bit-phase-flip noise, the approximation of $\eta=\infty$.
    }
    \label{fig:7}
\end{figure*}
\begin{figure}[!t]
    \centering
    \includegraphics[width=\linewidth]{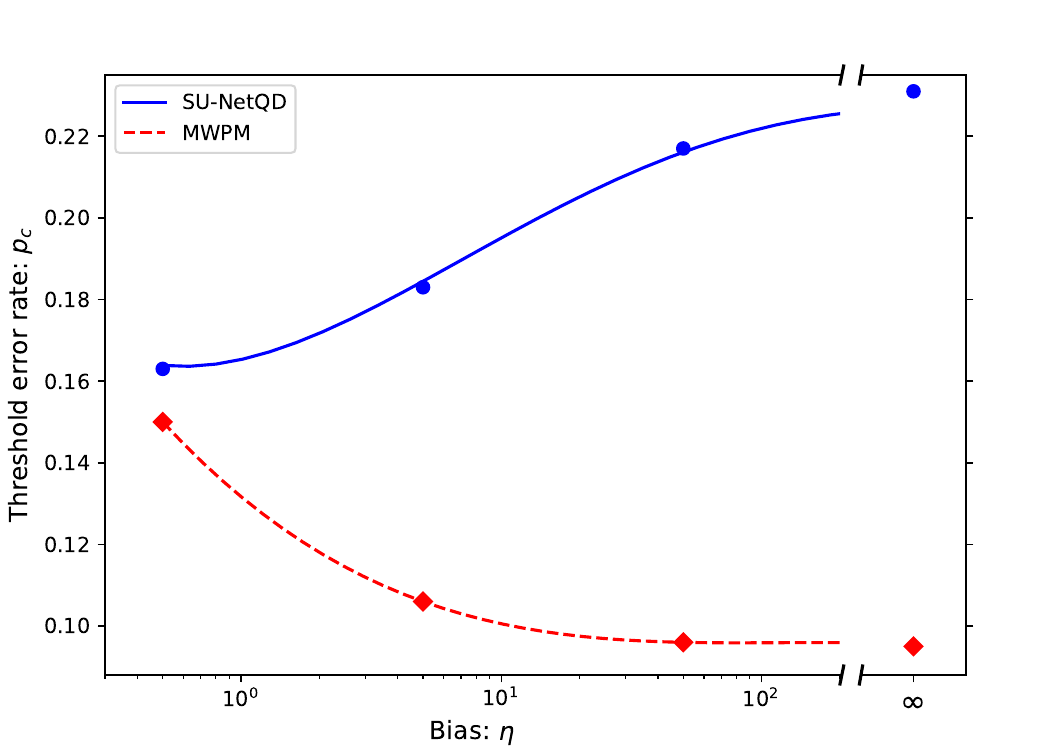}
    \caption{Threshold of error rate $p_c$ as a function of bias $\eta$.  The blue (red) points indicate the estimation of the threshold using the SU-NetQD (MWPM). The points at the largest bias value correspond to infinite bias $\eta$, which is equivalent to only Y error. The curves are the results of fitting the values using log-polynomial functions.}
    \label{fig:8}
\end{figure}

\section{Experiments} \label{sec:experiments}
In this section, we demonstrate our numerical simulations with our SU-NetQD. Specifically, we consider a type of biased noise model that covers depolarizing noise, bit-phase-flip, and other bias-level noise. We study the threshold of toric code under the biased noise and discover a high threshold of 0.231 for the extremely high biased noise case.
\subsection{Experimental scenarios}
We conducted our experiments of  toric codes with code distances of 5, 7, and 9, considering various biased noise models which include depolarizing noise, bit-phase-flip, and other bias-level noise.   
We opted not to separately study bit-flip and phase-flip noise because the MWPM algorithm is believed to be near-optimal for correcting such noise. Under each noise model, errors were randomly introduced independently and identically distributed (iid), and the resulting syndromes were extracted to form our dataset.

A Pauli error channel can be defined by an array \((1-p, p_{x}, p_{y}, p_{z})\), corresponding to the probabilities for each Pauli operator I (no error), X, Y, and Z, respectively. 
We introduce a bias parameter $\eta=\frac{p_y}{p_x+p_z}$, representing the ratio of the probability of a \(Y\) error to the total probability of non-\(Y\) Pauli errors. 
In our experiments, we only consider the case of \(p_{x}=p_{z}\) and study the effects of noise with different  $p_{y}$.  
With the physical error rate  \(p=p_{x}+p_{y}+p_{z}\), the probability of \(Y\) error  is \(p_{y}=\frac{\eta}{\eta+1}p\), and the probabilities of \(X\) and \(Z\) errors  are \(p_{x}=p_{z}=\frac{1}{2(\eta+1)}p\). 
With \(\eta=0.5\), the probability of each Pauli error is \(p/3\), corresponding to the depolarizing noise.  With \(\eta\to \infty\), the probability of a \(Y\) error approximates to \(p\), corresponding to the bit-phase-flip noise. With a non-zero $\eta$, the noise modes is noted as biased depolarizing noise. In our experiments, we study the case of $\eta=5$ and $\eta=50$ in the biased depolarizing noise.

In order to demonstrate the practicality of our decoder, we further investigate the performance of our decoder for toric codes in circuit-level noise where we include the measurement errors further in the depolarizing noise model mentioned above. The measurement noise is set with a probability $p_\text{m}=p$ in a custom way, we apply an X error before each measurement~\cite{varbanov2023neural}.

For the training of our SU-NetQD, the data label is set in the following way. When the MWPM method successfully decodes the quantum state and the number of single-qubit operations in the suggested recovery operation is less than the number of single-qubit errors that occurred, we set the label to be the recovery chain predicted by the MWPM decoder. Otherwise, we set the label to be the set of the original errors, which can be used as a recovery chain operation.

\begin{figure*}[!t]
    \centering
    {{\subfigure(a)}
    \includegraphics[width=0.46\linewidth]{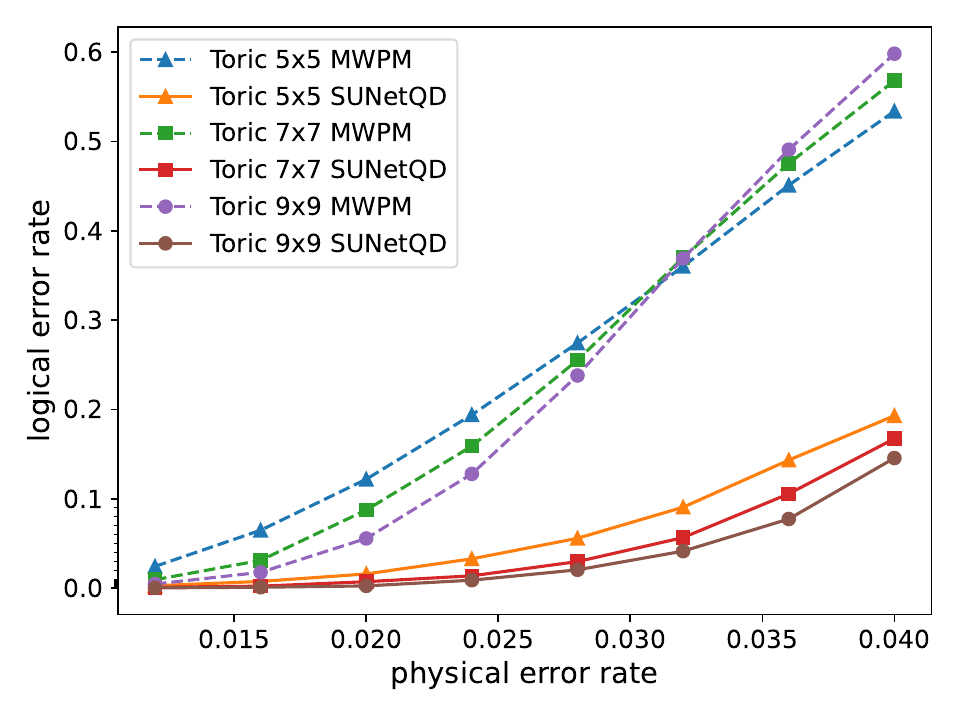} }
    {{\subfigure(b)}
     \includegraphics[width=0.46\linewidth]{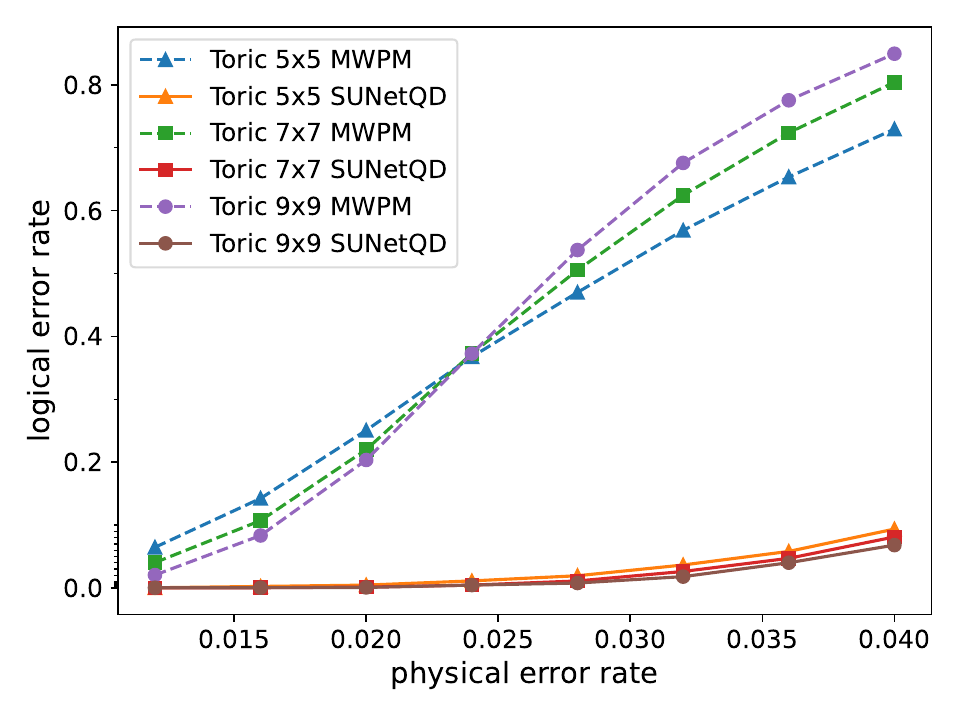}}
     {{\subfigure(c)}
     \includegraphics[width=0.46\linewidth]{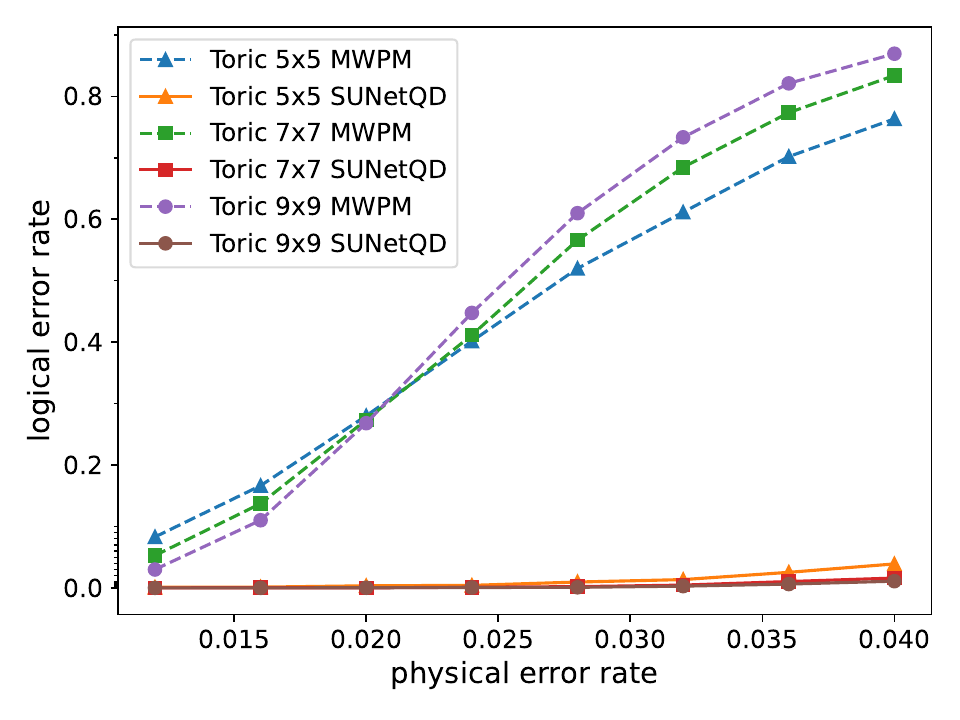} }
     {{\subfigure(d)}
   \includegraphics[width=0.46\linewidth]{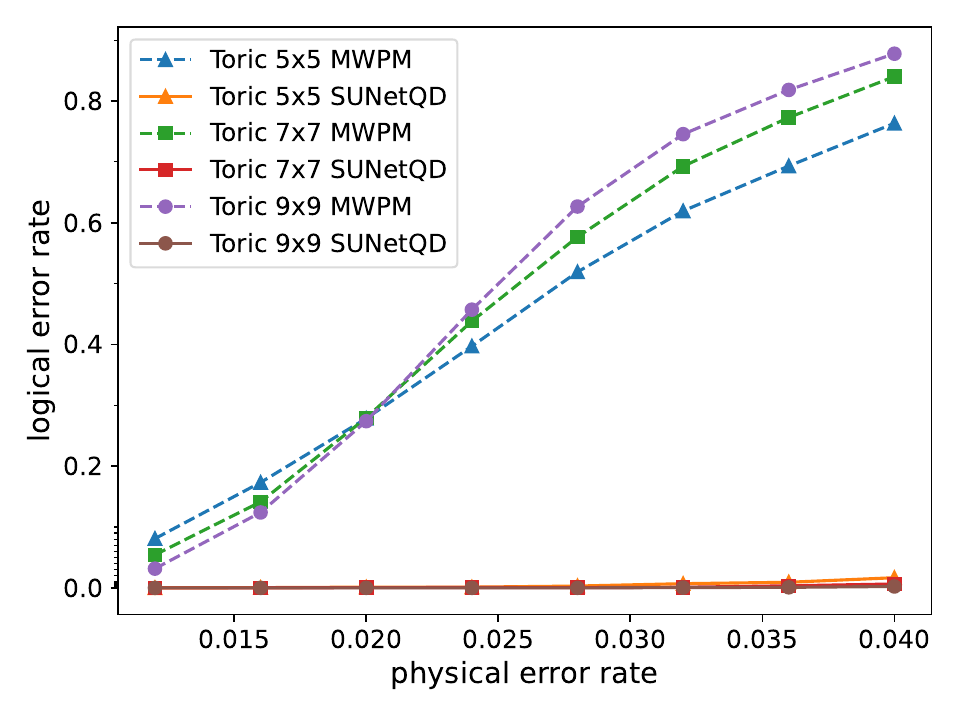} }
    \caption{Comparison of the decoding performance (logical error rate (LER)) of SU-NetQD and MWPM under measurement noise condition and various physical error rates of different noise models. 
   The X-axis represents the physical error rate, and the Y-axis represents the logical error rate of the decoding process. (a) depolarizing noise, 
   (b) biased depolarizing noise with $\eta=5$, 
   (c) biased depolarizing noise with $\eta=50$, 
   (d) bit-phase-flip noise, the approximation of $\eta=\infty$.}
    
    \label{fig:circuit-level-noise}
\end{figure*}

\subsection{Experimental setting}
The network structure of our SU-NetQD varies for the codes with different code distances. Specifically, we adapt the sampling layers in our low-level decoder for different code distances to ensure efficiency. For toric codes with code distances of 7 and 9, we employed sampling layers with channel numbers 16, 32, and 64. In contrast, for code distances of 5, we utilized only two sampling layers with channel numbers of 32 and 64. Within the model architecture, the self-attention block featured 4 heads, and the convolutional kernel size in the residual block was set to 3. The low-level decoder comprised a total number of 320k parameters, while the high-level decoder reused the structures of three low-level decoders, containing a total number of 980k parameters.

For bit-phase-flip noise, we train the two low-level decoders with error rates of 0.09 and 0.05 separately to learn the corresponding recovery chain predictions and then train the high-level decoder at an error rate of 0.13. 
Similarly, for the other biased noise models, we trained the two low-level decoders at error rates of 0.13 and 0.09 separately and the high-level decoder at an error rate of 0.17.

For a given code distance, we generated a dataset of 200k samples under the selected noise model and error rate. Model optimization was conducted over 50 epochs using the Adam optimizer, with a learning rate adjustment strategy starting from 0.01 and a weight decay strategy with a $\lambda$ value of 0.0001. The cross-entropy loss was utilized to classify recovery operations in the low-level decoder, and the binary cross-entropy loss was employed for the multi-label classification of logical errors in the high-level decoder. 

We evaluated the model over a range of physical error rates from 0.01 to 0.25 and, for each error rate scenario, generated a test set of 20k samples. The decoding process is considered successful when the recovery chains predicted by the low-level decoder completely eradicate the syndromes and the logical errors predicted by the high-level decoder match the joint effect of the recovery chains and the original errors. In our simulations, we use the logical error rate (LER) to evaluate the performance of our decoder, which is the ratio between the successful decoding cases and the total experimental cases.

The quality of the training dataset significantly influences the LER of decoder predictions, particularly for toric codes with small code distances. To leverage transfer learning, our decoder was initially trained from scratch on a dataset with a code distance of 15.  We implemented our models using PyTorch \cite{paszke2019pytorch} and qecsim \cite{qecsim}. All experiments were conducted on a single NVIDIA GeForce RTX 4090 GPU with 24GB of memory. The codes for our experiments are available through the GitHub link in \cite{Git}.

\subsection{Experimental results}
We evaluate the performance of our SU-NetQD by testing its decoder ability under depolarizing noise, biased depolarizing noise, and bit-phase-flip noise.  The results are shown in 
Fig.~\ref{fig:7}, where we compare the decoding performance of our SU-NetQD and the MWPM decoder across various code distances and physical error rates. 
Our experiments show that
our decoder consistently outperforms the MWPM decoder due to its ability to recognize correlations between different types of errors and adapt to varying error distributions effectively under different physical error rates. 
With an increase in the error rate, the decoding LER increases for both our SU-NetQD and MWPM. The rate of increase in our SU-NetQD is much lower than MWPM, which reflects the stability of our SU-NetQD performance. Under the same error rate for the same type of noise, the LER of our SU-NetQD is much lower than the LER obtained with MWPM.

In our experiments, although our decoders were trained on datasets with high error rates (e.g., \( p = 0.13 \) and \( p = 0.09 \)), they could be used in the low error rate cases and exhibit excellent decoding performance. For instance, at \( p = 0.01 \) and \( p = 0.05 \), the decoding LER of SU-NetQD surpasses that of MWPM. %This indicates that it does not tend to introduce additional errors in low-noise conditions. 
This performance is primarily attributed to our method of constructing dataset labels. We predominantly used the recovery chains predicted by the MWPM decoder, as they tend to be the shortest. Only when the MWPM decoder failed to decode or provided recovery operations that acted on more qubits than the error qubits, did we use the original error set as labels.

The threshold of a QEC under a certain noise environment is a key metric to evaluate the performance of a QEC. While the error rate is below the QEC threshold, the increase in code size can reduce the logical error rates. If the error rate is higher than the QEC threshold, the increase of code size is helpless for error correction and the information is not able to be corrected. 
The measure of the threshold for a QEC is an important task. Our SU-NetQD can obtain a more accurate QEC threshold compared to the MWPM, where the threshold discovered in our SU-NetQD is higher than the value discovered with MWPM as shown in Fig.~\ref{fig:7}. Our threshold of 0.163 for depolarizing noise is consistent with the known results~\cite{iolius2022performance,PhysRevX.6.041034}. The QEC threshold is obtained with the intersection of the decoding curves for the QEC with different code sizes. The decoding curves for different code sizes are ensured to intersect to reveal the QEC threshold~\cite{qecsim}. 

\begin{table*}[ht!]
\centering

\caption{Comparison of the logical error rate of MWPM with and without high-level decoder across different noise models and distances.}

\label{table:1}
\begin{tabular}{c|c|c|c|c|c|c|c|c|c|c}
\hline\hline
\multirow{2}{*}{Decoder} & \multirow{2}{*}{distance} & \multicolumn{3}{c|}{Bit-flip} & \multicolumn{3}{c|}{Depolarizing} & \multicolumn{3}{c}{Bit-phase-flip} \\ \cline{3-11} 
                         &                              & p=0.05   & p=0.09   & p=0.13   & p=0.05   & p=0.13   & p=0.21   & p=0.05   & p=0.13   & p=0.21   \\ \hline
\multirow{3}{*}{MWPM}    &  d=5                          & 3.30\% & 17.76\% & 30.14\% & 1.35\% & 28.82\% & 66.77\% & 5.60\% & 55.17\% & 86.875\% \\
                         & d=7                          & 1.54\%     & 16.58\%     & 43.77\%      & 0.45\% & 25.54\% & 72.7\%    &2.87\%     & 61.82\%    & 90.70\%     \\
                         & d=9                         & 
                         0.83\%& 15.33\% & 46.08\% & 0.2\% & 23.38\% &  77.26\%& 1.56\%&  66.03\%& 99.93\%\\ \hline
\multirow{3}{*}{enhanced\_MWPM} & d=5                         &  3.24\%  & 17.58\% & 38.63\% & 0.96\% & 23.73\%& 62.39\%&0.32\% & 19.69\% &64.14\%\\
                         & d=7                          & 1.49\% & 16.28\% & 42.86\%&  0.19\%& 19.4\% & 68.68\% &  0.005\%&  13.01\% &  67.45\%\\ 
                         & d=9                         & 0.79\%  & 15.15\% & 45.61\% & 0.05\% & 16.84\% & 72.42\% & 0.001\% & 9.11\% & 70.56\% \\ \hline\hline
\end{tabular}
\end{table*}

\begin{figure}[!t]
    \centering   \includegraphics[width=\linewidth]{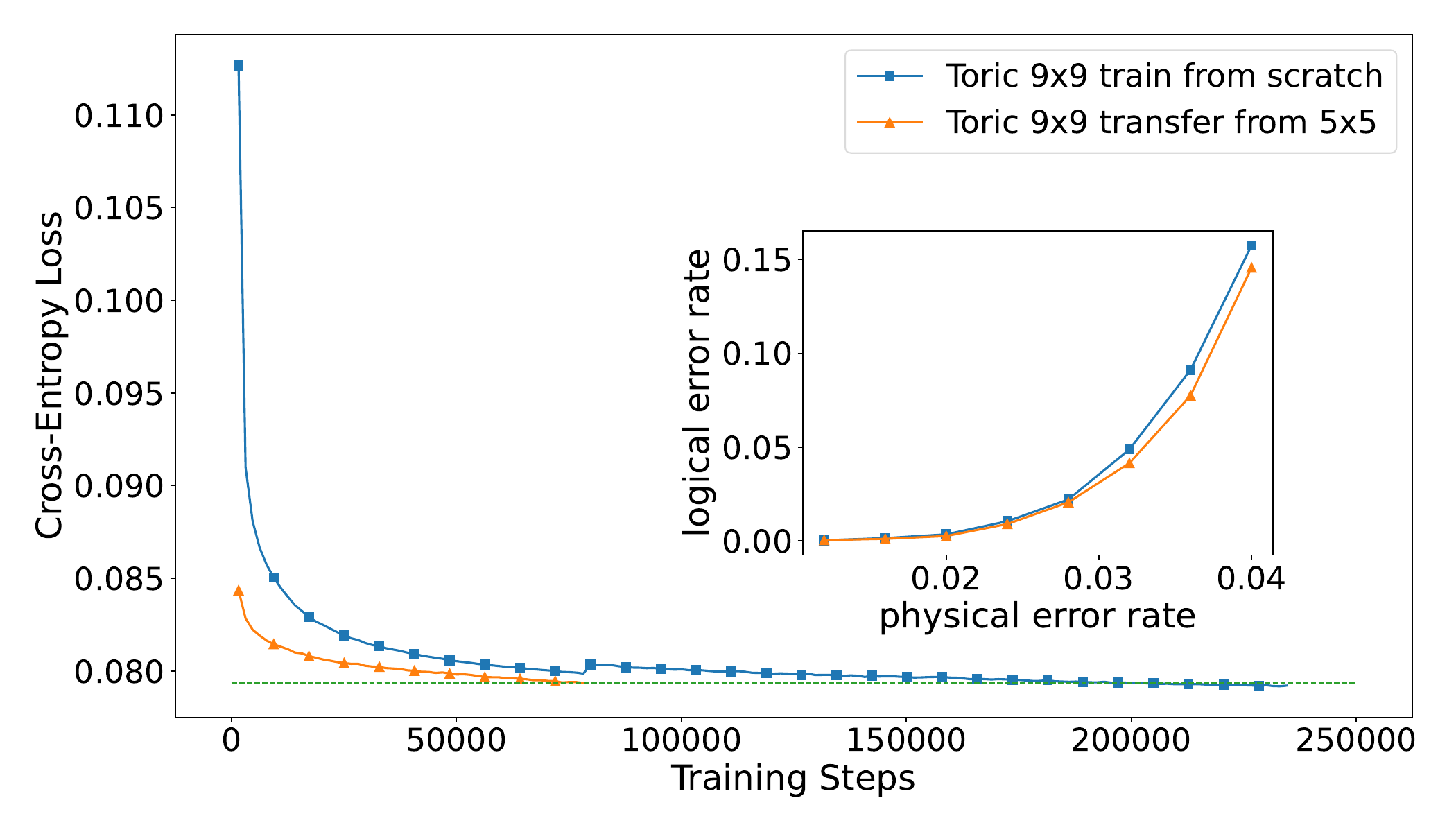}
    \caption{The comparison of training time required for our decoder and the defective versions without transfer learning mechanics.}
    \label{fig:compare-transferlearning}
\end{figure}

We discover an increased trend of the threshold for toric code in biased depolarizing noise with the increase of bias value $\eta$ using our SU-NetQD as shown in  Fig.~\ref{fig:8}. 
In the limit of \(\eta \to \infty\), i.e., under the bit-phase-flip noise channel, the threshold measured by our model is 0.231 compared with the threshold value of  0.095  revealed by MWPM. 
In contrast, Using MWPM we discover a decreasing trend of the threshold for toric code in biased depolarizing noise with the increase of bias value $\eta$ as shown in  Fig.~\ref{fig:8}, which resulted from the ability deficiency of MWPM in dealing with the correlated noise.

While considering the measurement noise, we investigate the performance of our method as shown in Fig.~\ref{fig:circuit-level-noise}. The logical error rates revealed with our SU-NetD are much lower than the discovered by MWPM. Also, the thresholds discovered by MWPM is much lower which does not reveal the real error correction ability of toric code.  Our decoder has the ability to decode the toric code at the region that MWPM is unable to do the error correction. 

Transfer learning enables the efficient training of our model as shown in Fig.~\ref{fig:compare-transferlearning}. For instance, while our network trained with cases with code size of $5\times5$ is transferred to the training of cases with $9\times9$ code distance, the time consumed by the former is 1/5 of the latter, and the LER is higher than the latter.

\begin{figure*}[!t]
    \centering {\subfigure(a)} \includegraphics[width=0.41\linewidth]{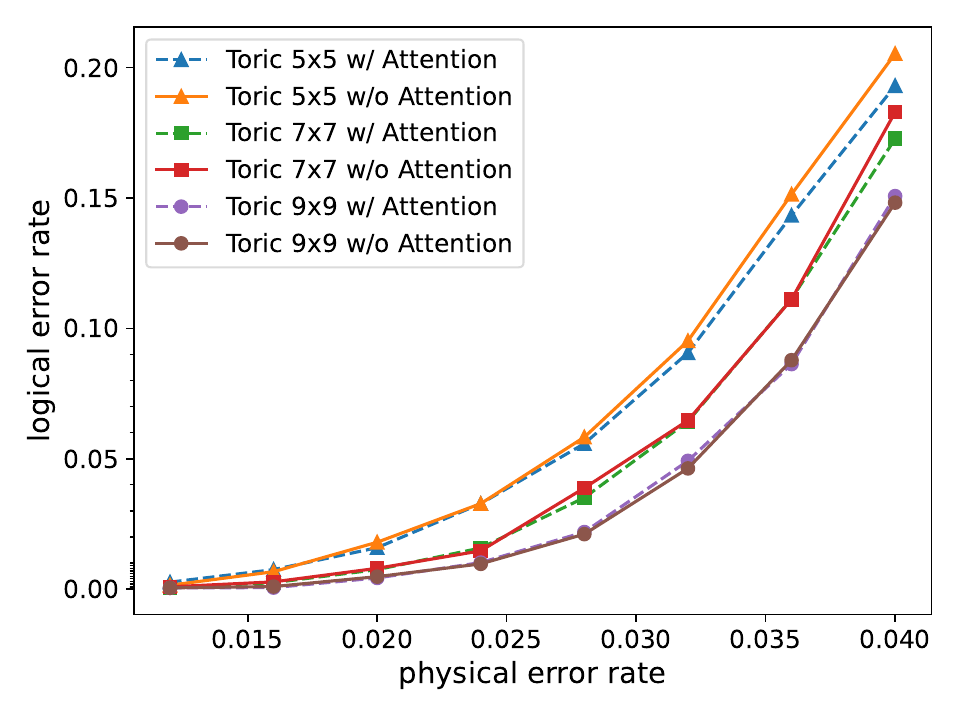}{\subfigure(b)}\includegraphics[width=0.48\linewidth]{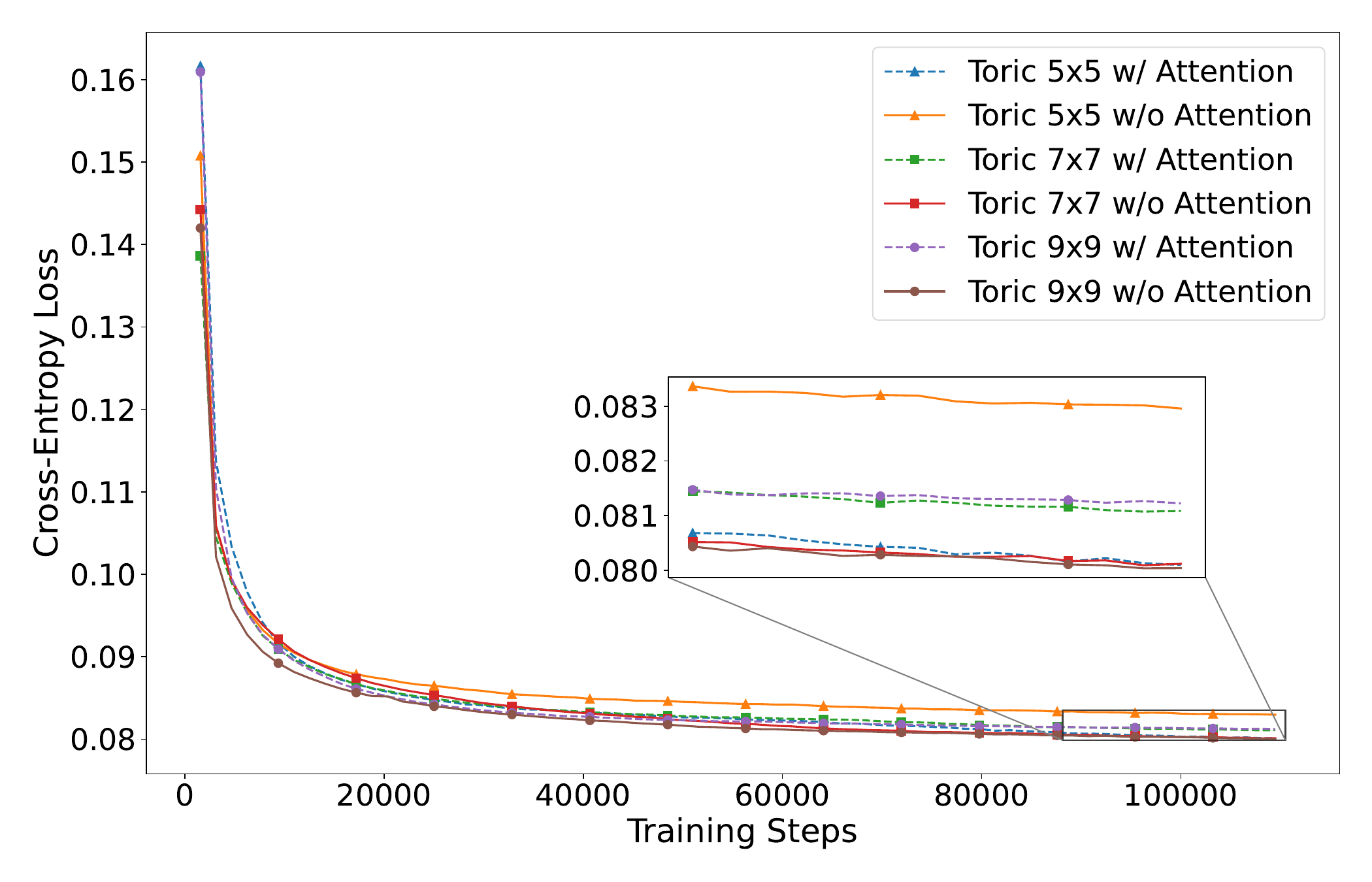}
    \caption{The comparisons between our decoder and the defective versions of our decoders without self-attention mechanics with various code size, (a) the logical error rates comparisons, (b) the loss curves in the training process. }
    \label{fig:compare-selfattention}
\end{figure*}

Self-attention enables our model to have the ability to identify the long-range error chain cases. We compare our model and the defective version of our model by removing the self-attention as shown in Fig.~\ref{fig:compare-selfattention}.  We find that with the self-attention mechanism, our method performs better in two cases, one is the case the physical error rate is high, the other is the code distance is small. We observe that with self-attention, as the physical error rate increases for small code distances, our decoding logical error rates get lower.

Furthermore, our decoder design is efficient and compatible with the current known decoder. In our experiments, we concatenate the MWPM decoders with our high-level decoder and investigate the performance enhancements. The high-level decoder helps identify and correct logical errors that MWPM may introduce during decoding, enhancing the overall accuracy.
Our experimental results are shown in Table \ref{table:1}, where the enhanced\_MWPM refers to the results of post-processing MWPM's decoding output using the high-level decoder. It is evident that the performance of MWPM often significantly improves after employing the high-level decoder. Even under the bit-flip noise model, the high-level decoder can enhance the decoding accuracy of MWPM a little, suggesting that while MWPM is nearly optimal, it does not achieve theoretical optimality. 
These findings highlight the effectiveness of integrating advanced neural network-based strategies in our decoder for complex noise environments where traditional methods like MWPM may fall short, such as the integration of our high-level decoder in quantum error correction schemes.

\section{Conclusions} \label{sec:conclusions}
In this work, we propose a scalable efficient quantum toric code decoder that uses a self-attention U-Net decoder structure. Our SU-NetQD composite of low-level and high-level decoders finely handles the error correction processes, which outperforms the MWPM, especially in the circuit level noise cases. With our SU-NetQD, we discover an increased trend of the thresholds for the increased biased depolarizing noise and a threshold value of 0.231 while $\eta=\infty$. Our decoder provides a practical tool for the study of QEC and promotes the practical application of quantum error correction codes.

\begin{acknowledgments}
We appreciate the discussion with Dr. Benjamin J. Brown. We acknowledge support from the National Natural Science Foundation of China (Grant No.  12104101) and the Fundamental Research Funds for the Central Universities, Stability Program of National Key Laboratory of Security Communication (2023), the Major Research Project of National Natural Science Foundation of China under Grant 92267110, the Joint Funds of the National Natural Science Foundation of China (Grant No. U22B2025) and the Key Research and Development Program of Shaanxi  (Grant No. 2023-GHZD-42).
\end{acknowledgments}

\bibliography{reference}

\end{document}